\documentclass[conference]{IEEEtran}
\IEEEoverridecommandlockouts
\usepackage{cite}
\usepackage{amsmath,amssymb,amsfonts}
\usepackage{algorithmic}
\usepackage{graphicx}
\usepackage{textcomp}
\usepackage{xcolor}
\usepackage{array}
\usepackage{hyperref}
\def\BibTeX{{\rm B\kern-.05em{\sc i\kern-.025em b}\kern-.08em
    T\kern-.1667em\lower.7ex\hbox{E}\kern-.125emX}}

\usepackage{url}

\begin{document}

\onecolumn
\thispagestyle{empty}

\vspace*{1cm}
{\Huge \noindent IEEE Copyright Notice \par}
\vspace{1.2cm}

\noindent
\setlength{\fboxsep}{18pt}%
\setlength{\fboxrule}{0.6pt}%
\fbox{%
  \begin{minipage}{\dimexpr\linewidth-2\fboxsep-2\fboxrule\relax}%
    \large
    \copyright~2023 IEEE. Personal use of this material is permitted. Permission from IEEE must be obtained for all other uses, in any current or future media, including reprinting/republishing this material for advertising or promotional purposes, creating new collective works, for resale or redistribution to servers or lists, or reuse of any copyrighted component of this work in other works.\\[2em]

    This is the author's accepted version of a paper published in: IECON 2023- 49th Annual Conference of the IEEE Industrial Electronics Society.\\[2em]

    DOI: \href{https://doi.org/10.1109/IECON51785.2023.10312235}{\texttt{10.1109/IECON51785.2023.10312235}}%
  \end{minipage}%
}%

\vfill
\clearpage
\twocolumn

\title{Analytics for the Optimization of the Soybean Oil Purification Process\\}
\author{\IEEEauthorblockN{Henrik Meyer}
\IEEEauthorblockA{\textit{Faculty of Technology} \\
\textit{University of Applied Sciences Emden/Leer}\\
Emden, Germany \\
henrik.meyer.1@stud.hs-emden-leer.de}
\and
\IEEEauthorblockN{Lars Ahlers}
\IEEEauthorblockA{\textit{Faculty of Technology} \\
\textit{University of Applied Sciences Emden/Leer}\\
Emden, Germany \\
lars.ahlers@stud.hs-emden-leer.de}
\and
\IEEEauthorblockN{Pedro Querini, Erica Fernandez}
\IEEEauthorblockA{\textit{Industrial Engineering} \\
\textit{UTN - FRRA, Rafaela}\\
Rafaela, Argentina \\
(pedro.querini, erica.fernandez)\\@frra.utn.edu.ar}
\and
\IEEEauthorblockN{Maria L. Caliusco}
\IEEEauthorblockA{\textit{Industrial Engineering} \\
\textit{UTN - FRSF, Santa Fe}\\
Santa Fe, Argentina \\
mcaliusc@frsf.utn.edu.ar}
\and
\IEEEauthorblockN{Martín A. Bär}
\IEEEauthorblockA{\textit{Faculty of Technology} \\
\textit{University of Applied Sciences Emden/Leer}\\
Emden, Germany \\
martin.baer@hs-emden-leer.de}
\and
\IEEEauthorblockN{Armando W. Colombo [FIEEE]}
\IEEEauthorblockA{\textit{Faculty of Technology} \\
\textit{University of Applied Sciences Emden/Leer}\\
Emden, Germany \\
awcolombo@ieee.org}

}

\maketitle

\begin{abstract}
Machine Learning, Artificial Intelligence, among others, are very promising methodologies and technologies that are emerging for implementing a broad spectrum of analytics within digitalized eco-systems. Analytics containing adequate analytical models are generating a burgeoning interest from Business Intelligence-, Information Technology (IT)- and Operational Technology (OT) -professionals, who are able to exploit the huge amount of internally and externally available data and information that lies behind digitalized components and systems and their associated processes.

In this paper, the authors present the essential specifications of an analytical model developed and implemented applying the Knowldege Discovery in Databases (KDD) approach. The analytical model is the essential part of an analytics component, positioned as an digitalized asset within an Industry 4.0-compliant (RAMI 4.0) infrastructure, and  used to optimize the industrial Soybean Oil Purification Process associated to the digitalized eco-system. \newline
\end{abstract}

\begin{IEEEkeywords}
Analytics, Analytical Model, KDD, Machine Learning, Regresion Trees, Digitalization, Soybean Oil Prurification Process
\end{IEEEkeywords}

\section{Introduction}
A brief screening of current published reports allows to identify a huge amount of work being done to simulate technical processes for doing process prediction or optimization, applying analytical models, based on machine learning methods \cite{IBM.2022, Deloitte.2020}.
In many cases, the data needed is already available due to automation and the use of sensors and Programmable Logic Controllers (PLCs) in the course of Industry 3.0 \cite{Jiang.2022}. In the sense of Industry 4.0, the physical and digital worlds can be linked here to generate added value by using the available data \cite{DiezOlivan.2019}. \newline

Several papers already exist in which analytical models based on machine learning have been developed and implemented. The goal is usually to simulate the technical process, based on the process data and then to use this simulation for process prediction or optimization \cite{ Weichert.2019}.

Successful examples of implementations can be seen in various applications \cite{Sumayli.2023, Gim.2023, Masinelli.2020}. In a study on electrical discharge machining, an artificial neural network was used to model the process with regard to the selected parameters \cite{Ugrasen.2014}. Good process predictions were reached, resulting in the creation of a model that allows for optimization of the process parameters.

In another study, different machine learning models were compared predicting process parameters in the production of biodiesel from soybean oil \cite{Sun.2022}. A multi-layer perceptron (neural network type), regression tree and k-nearest-neighbour were created. Multi-layer perceptron and k-nearest-neighbor performed better than the regression tree and attained a satisfactory level of predictive accuracy.

Despite the increase in investments, such systems are still an exception rather than the rule, especially in small companies \cite{IBM.2022, Deloitte.2020}. The reason for this is often a lack of personnel, resources and tools for implementation \cite{businesswire.2019}. \newline

In this paper, the authors present the essential specifications of an analytical model developed and implemented applying the Knowldege Discovery in Databases (KDD) approach. The analytical model is an essential part of an analytics component, positioned as a digitalized asset within an Industry 4.0-compliant (RAMI 4.0) infrastructure. It is used to minimize losses in the purification of soybean oil and in consequence for optimizing the industrial soybean oil process as a whole.

A key requirement for the design of the analytics is the ability to implement the associated analytics using low cost hardware and possibly free open source software to maximise the business impact of the solution. \newline

Following this introduction, which includes a brief state-of-the-art analysis, section II of the paper describes the major specifications and technical characteristics of the soybean oil purification process. After that, in section III, the analytics is classified within a standard reference architecture (RAMI 4.0) and the specification of the analytical model is worked out in section IV. Section V briefly presents the evaluation and optimization possibilities and in the last section a conclusion is given.

\section{Description of the Soybean Oil Purification Process}

\begin{figure}[htbp]
\centerline{\includegraphics[width = 0.95\columnwidth]
{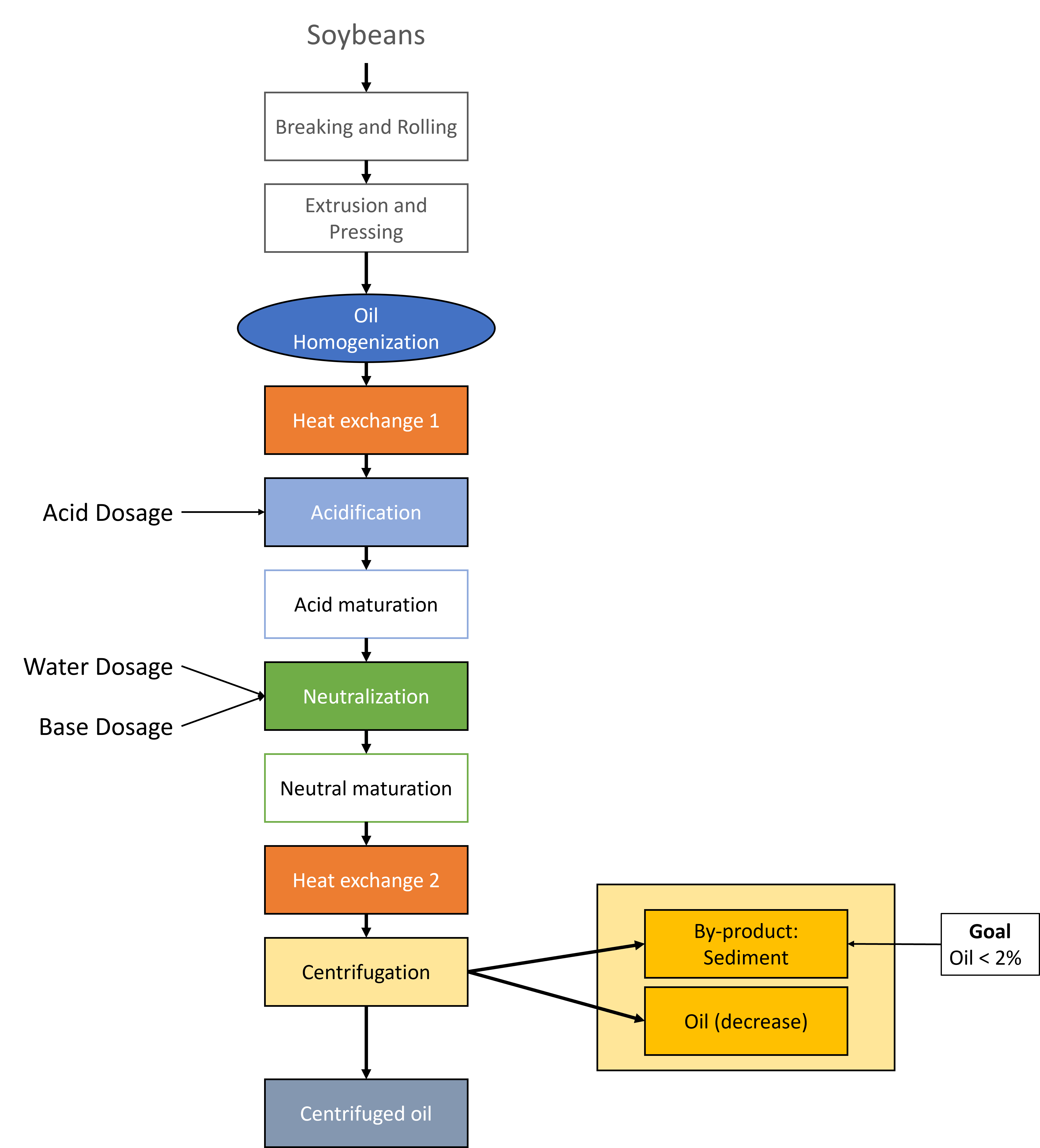}}
\caption{Process of soybean oil purification - upstream process steps in gray}
\label{fig: Process Diagram}
\end{figure}

The production of oil from soybeans starts with harvesting. From the soybeans, the crude oil is extracted. The most important steps towards the crude oil are crushing and rolling, followed by extrusion and pressing of the beans \cite{OBrien.2008}.
After these steps, impurities in the crude oil have to be removed. This process, along with the preliminary steps, is shown in Figure \ref{fig: Process Diagram}.

The first step after homogenization is to heat the crude oil to a specific temperature. Acid is added to the oil. Acid maturation follows, during which the impurities are dissolved by the acid. In order to bring the oil back to a neutral pH (potentia hydrogenii) and thus ensure the quality of it, base is added. Additionally, water is added to help with removing the impurities. In the following maturation phase, the base neutralizes the acid. After that, the oil should have a neutral pH. Finally, the oil mixture is cooled down. With the help of centrifuges, the oil is separated from the remaining sediment.

The sediment that is separated from the actual oil in the final process step still contains a certain amount of oil that can no longer be used.
The aim is to minimize this amount of oil in the sediment. According to the company, where this process is analyzed, the oil content should be under 2 \% (in relation to the amount of incoming crude oil), which is currently only reached irregularly.

During the process, several measurements are taken, containing selected parameters such as the acid flow, given parameters such as the pH of the crude oil and the outcome represented by the obtained oil, the sediment and the final pH.
Currently, the process parameters are chosen by performing sample tests. These sample tests may not keep up with the changes of given parameters influencing the process. An analytical model trained with available data from the process could help by calculating optimal parameters in dependence of the given process parameters.

\section{Placement of the Analytics within a Standard Reference Architecture}

\begin{figure}[htbp]
\centerline{\includegraphics[width = 0.9\columnwidth]
{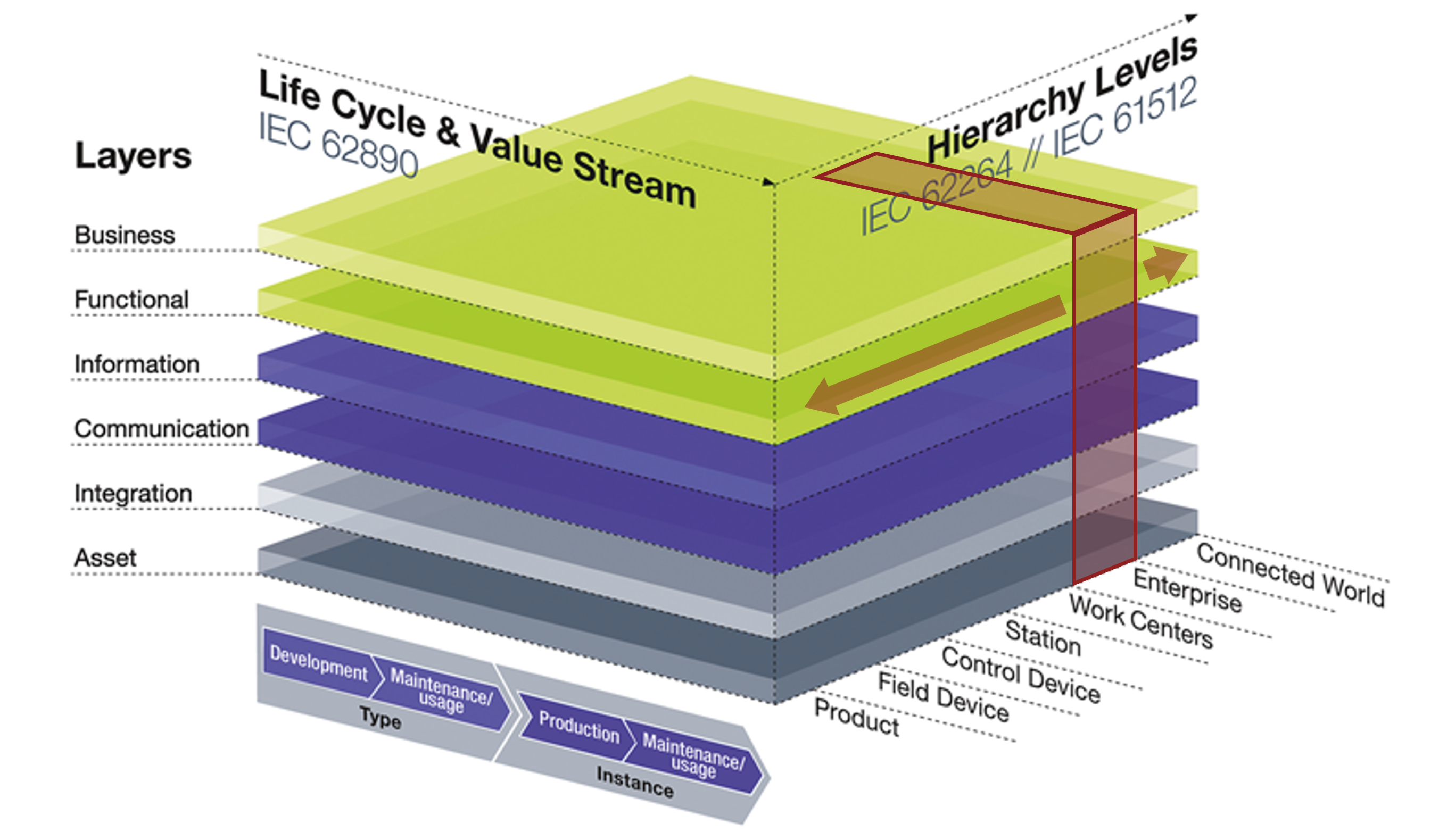}}
\caption{Position of the analytics within RAMI 4.0 \cite{Heidel.2017}}
\label{fig: RAMI 4.0}
\end{figure}

The analytics to be developed can be assigned to the area of Industry 4.0. The placement of the analytics in the reference architecture model helps to identify (i) which phases of the analytics life cycle are being considered and (ii) the set of interactions between the analytics and other digitalized components and systems within the companies infrastructure. As an established model standardized in DIN SPEC 91345 \cite{Din91345.2016}, the Reference Architecture Model for Industry 4.0 (RAMI 4.0) is suitable for this classification, as shown in Figure \ref{fig: RAMI 4.0} \cite{Heidel.2017}.

The analytics containing the analytical model is a component positioned within the enterprise level (e.g. Level 3 following the IEC 62264 \cite{Iec62264.2013}). As a consequence and following the specifications of RAMI 4.0, it will act as another digitalized asset, able to interact with any other asset of the digitalized eco-system, from its position in the enterprise level. The arrows in Figure \ref{fig: RAMI 4.0} illustrate these expected interactions with other hierarchy levels. The analytics is an asset, that has an interoperability and communication feature (e.g. OPC UA, MQTT) allowing it to communicate within the network. With the help of this communication capability, it can gather data from the process and offer its function of predicting optimal process parameters to support, among others, the business goals of the company (improving the decision-making process).

Since the analytics is working on the operation of the plant, it is classified as an instance according to the IEC 62890 \cite{Iec62890.2017}, covering both sub-phases.

\subsection{Placement of the Analytics within the automated and digitalized Infrastructure}

\begin{figure}[htbp]
\centerline{\includegraphics[width = 0.98 \columnwidth]
{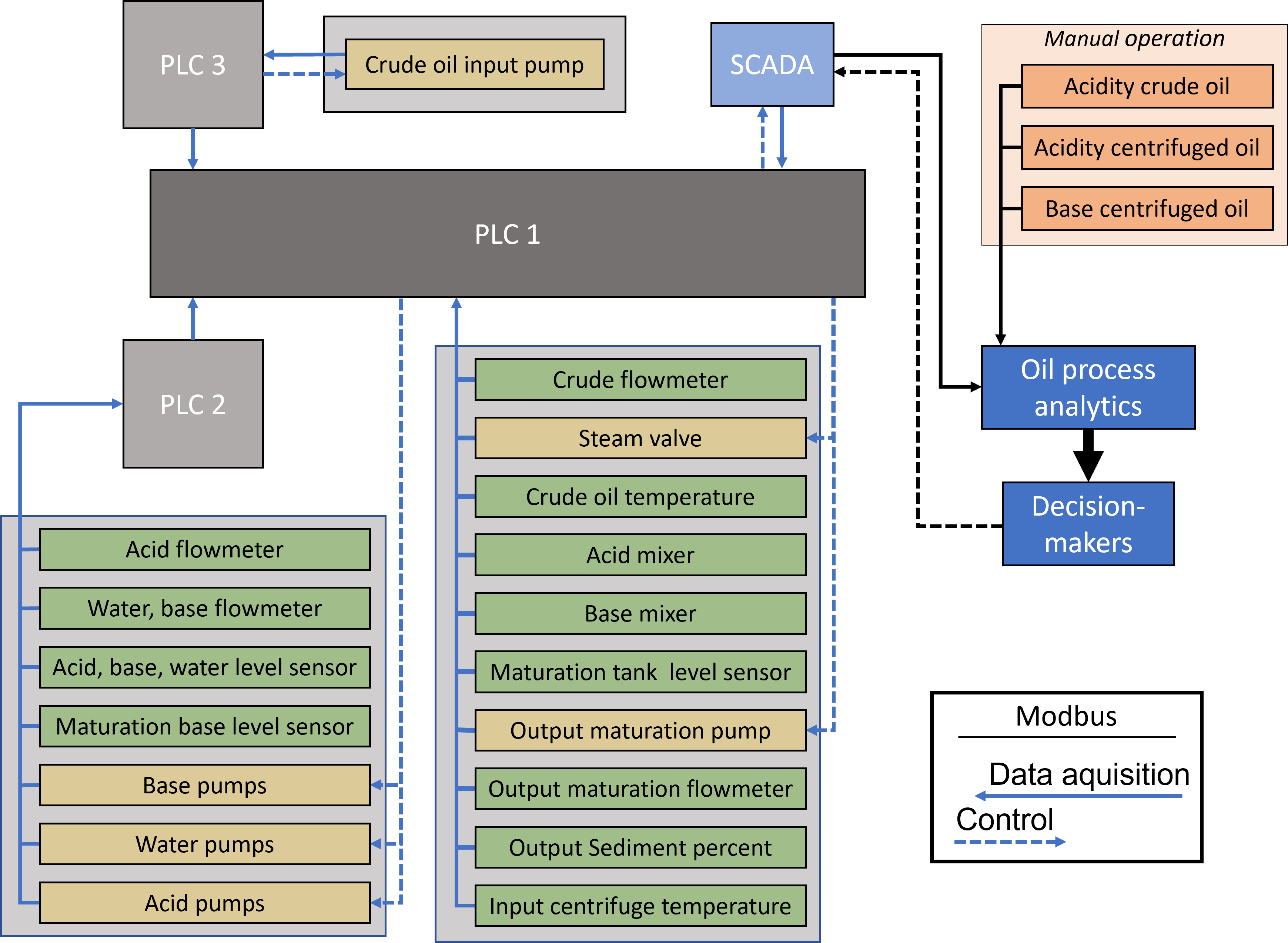}}
\caption{Placement of analytics within automated and digitalized infrastructure}
\label{fig: Process Infrastructure}
\end{figure}

One aspect that is particularly important in the integration and communication layers, is the placement of the analytics within the existing infrastructure.
Figure \ref{fig: Process Infrastructure} shows this placement and the connection to other components. \newline

Components like sensors or actuators at the field device level send data to the different PLCs in the control device level and/or get controlled by them. From there, the data is sent to a Supervisory Control and Data Acquisition (SCADA) system. The communication between the industrial devices is realized using Modbus.
The analytics to be developed is connected to this SCADA, receiving the selected data. Additionally, manually collected data is received.
Based on the data, the supposed functionality of the analytics is carried out. The results are then transferred to the decisionmakers, controlling the process using the SCADA system.

It should be noted here that other systems beyond the decisionmakers can receive the results of the analytics. Such a function can be implemented, for example, by offering the results deploying an OPC UA server. Different clients can then subscribe to the information as needed. In addition to OPC UA, other Industry 4.0 compatible communication technologies such as MQTT can be used \cite{Heidel.2017}.

\section{Knowledge Discovery in Databases Process}

With the aim of ensuring a structured approach specifying the analytics, the process is based on the KDD approach.

\subsection{Database}

The data needed to build the analytical model is available through a SCADA system, filled by PLCs and in the form of spreadsheet lists filled by the personnel. The values in the spreadsheet lists are determined in chemical experiments, with one result being the acidity of the crude oil \cite{InternationalOrganizationforStandardization.2020}. All data measured in the process is shown in Table \ref{tab: All data in process}.

\begin{table}[htbp]
\caption{\emph{Data measured during the oil purification process - phases are analogous to the phases shown in Figure \ref{fig: Process Diagram}}}
\label{tab: All data in process}
\begin{center}
\begin{tabular}{ | p {0.4cm} | p {2cm} | p {3cm} | p {0.5cm} | p {0.6cm} | }
	\hline
	\textbf{No.} & \textbf{Phase} & \textbf{Description} & \textbf{Unit} & \textbf{Freq.} \\
	\hline 1 & Initial Condition & Acidity (crude oil) & \% & 4 h \\ \cline{1-1} \cline{3-5}
	2 &  & Flow crude oil & Kg/h & 10 s \\ \cline{1-5}
	\hline 3 & Heat exchange 1 & Temperature & $^{\circ}$C & 10 s \\ \cline{1-5}
	\hline 4 & Acidification & Flow acid & l/h & 10 s \\ \cline{1-5}
	\hline 5 & Acid maturation & Acidity & \% & 4 h \\ \cline{1-5}
	\hline 6 & Neutralization & Flow base & l/h & 10 s \\ \cline{1-1} \cline{3-5}
	7 &  & Flow water & l/h & 10 s \\ \cline{1-5}
	\hline 8 & Neutral maturation & Residence time & s & 10 s \\ \cline{1-5}
	\hline 9 & Heat exchange 2 & Temperature & $^{\circ}$C & 10 s \\ \cline{1-5}
	\hline 10 & Centrifugation & Flow (Centrifuge IN) & Kg/h & 10 s \\ \cline{1-1} \cline{3-5}
	11 & & Flow (Centrifuge OUT) & Kg/h & 10 s \\ \cline{1-1} \cline{3-5}
	12 & & Oil content sediment & \% & 10 s \\ \cline{1-1} \cline{3-5}
	\hline 13 & Final condition & Acidity (centrifuged oil) & \% & 1 h \\ \cline{1-1} \cline{3-5}
	14 & & Base (centrifuged oil) & ppm & 1 h \\ \hline
\end{tabular}
\end{center}
\end{table}

All data should be available in a database so that they can be used for training the model and possibly also for use in other projects. Both the data from the spreadsheets and the PLC data must be automatically transferred to the database during later operation. For the transmission of the PLC data, for example, the OPC UA server functionality of the SCADA can be used. With an OPC UA client, the data can then be received and stored in a database.

\subsection{Machine Learning Model Selection}

Once it is known which data are fundamentally available, a suitable machine learning model can be preliminary selected. For the described task, in which a simulation of the process is to be created in order to optimally select certain parameters, various models can be used. One important aspect in the selection is the comprehensibility of the model for the users who will work with the analytics after completion.

Some of the most appropriate models are neural networks, multiple linear regression and decision trees.
In this case, the decision tree or regression tree seems to be the optimal model. In comparison to the neuronal network, it is much more intuitive (a neural network is like a black box), the regression tree is easier to train, it needs less data preparation and it is often easier to optimize \cite{Westreich.2010}. In fact, the results of a decision tree can usually be displayed graphically. This allows any user inspecting the results to easily interpret them with minimal training.

The multiple linear regression model could be a similarly good approach to the problem at hand. However, if more than two predictors are needed for the prediction, a major advantage of linear regression, the simple representation, is mostly lost.

When talking about regression trees, the random forest is a widely used algorithm in practice. In this case it is not used, because a clear representation is hardly possible \cite{Ziegler.2014}.

For creating and training the analytical model, the widely used library scikit-learn (sklearn) for Python will be used. One advantage of this library is that it can also be executed using a single-board computer like the Raspberry Pi.
This means that this analytical model can be created and trained using these types of devices that can be embedded with other assets.

\subsection{Data Selection}

Of all the available process data, shown in Table \ref{tab: All data in process}, those must first be selected that decisively describe the process and are as independent as possible of other variables. These variables then become the predictors or target variables of the subsequent analytical model. On one hand, there should be enough predictors to describe the process, on the other hand, too many can also have disadvantages. These disadvantages include overfitting, where the model adapts too much to the training data and no longer performs well in reality. Another possible disadvantage can come from redundancies, where aspects of the process are weighted disproportionately \cite{Ying.2019}.

The selection of predictors can be driven both based on knowledge about the process, and/or based on the data itself. Because of legal restrictions, mainly a knowledge and logic driven selection takes place.

All data presented will be gone through and an assessment will be given in each case. The numbers (x) shown below refer to the numbers used in Table \ref{tab: All data in process}. \newline

\begin{itemize}
	\item Initial condition - Acidity (1) and flow (2) of crude oil: Both parameters are the only ones measured in the process that cannot be actively influenced. They describe the oil entering the process and thus define how the following parameters should look. They are not dependent on other parameters considered in this process and their value cannot be considered as constant. Both parameters need to be included in the analytical model.
	\item Heat exchange 1, 2 - Temperature (3, 9): Both parameters are directly set and thus highly consistent. For this reason, a change could affect the outcome of the process. However, since both values always remain nearly constant, there is no need to consider them in a model and because of the lack of variance in the data, they are additionally difficult to model.
	\item Acidification - Flow of acid (4): The added acid does the main work in the process. The amount selected is a direct response to the acidity (1) and flow (2) of the crude oil being processed. It is neither constant nor dependent on other parameters.
	\item Acid maturation - Acidity (5): This parameter is directly dependent on the initial flow of crude oil (1, 2) and the flow of acid (4).
	\item Neutralization - Flow of base (6): The flow of base is a direct response to the amount of acid in the oil at this point of the process. It must be perfectly adapted to it to ensure a neutral pH and base of the end product. It is neither constant nor dependent on other parameters.
	\item Neutralization - Flow of water (7): It is not dependent on other variables or constant and has a potential influence on the outcome of the process.
	\item Neutral maturation - Residence time (8): The residence time behaves similar to the temperature (3, 9), it is configured directly and is highly constant.
	\item Centrifugation - Flow IN (10) and OUT (11): The volume flow into the centrifuges is directly dependent on the flow of crude oil (1), acid (4), base (6) and water (7). The difference between the volume flow into the centrifuges and out of the centrifuges describes the flow of sediment. This variable is again heavily dependent of the difference between the flow into the centrifuges and the flow of acid, base and water and the percentage of oil contained in the sediment (12). Furthermore, the variables are a result of the process, using them as a predictor is impracticable.
	\item Centrifugation - Flow of oil in the sediment (12): The percentage of oil contained in the sediment is the actual reason for the development of the analytical model or, in other words, the target variable.
	\item Final condition - Acidity (13) and base (14) in the centrifuged oil: The acid and base content of the centrifuged oil is an important criterion for its quality. They are very important for quality control, but also highly dependent on the ratio of acid in the crude oil (1, 2) and the flow of acid (4) and base (6).
\end{itemize}

Based on the assessment, the acidity of the crude oil (1), the flow of crude oil (2), acid (4), base (6) and water (7) and the oil content in the sediment (12) remain as important process parameters. The flow of oil in the sediment is the target variable, all other parameters are predictors. \newline

The selection of parameters made so far relies on the oil purification process being understood and logical relationships being correctly made. In order to substantiate the selection, a data-driven analysis would have to be carried out. This could both test theories and reveal new relationships not previously considered.
There are many different data driven methods to test whether there are correlations between two or more variables and how significant they are. One of those methods is linear (or polynomial) regression, where a straight line or higher order linear function is fitted to the measured values.
With the help of the model created, the correlation between the variables can then be calculated (e.g. Pearson’s correlation coefficient) \cite{Field.2018}. The different outcomes are compared, to state how strong relations between the different predictors are and how well they describe the target variable \cite{Brandt.2013}.

\subsection{Data Preparation}

Since the creation of the analytical model with the library sklearn is performed in Python, pandas and numpy are suitable libraries to include for processing and analyzing the data.
The data can first be loaded from the database with an SQL query and then this data can be transformed to a pandas dataframe.
The timestamp is used as a criterion for selecting the data imported from the database. As a first approach, the period of the last five working days is proposed. With five days, twenty four hours per day and data available every ten seconds, that's 43,200 rows of data.

\begin{figure}[htbp]
\centerline{\includegraphics[width =0.6\columnwidth]
{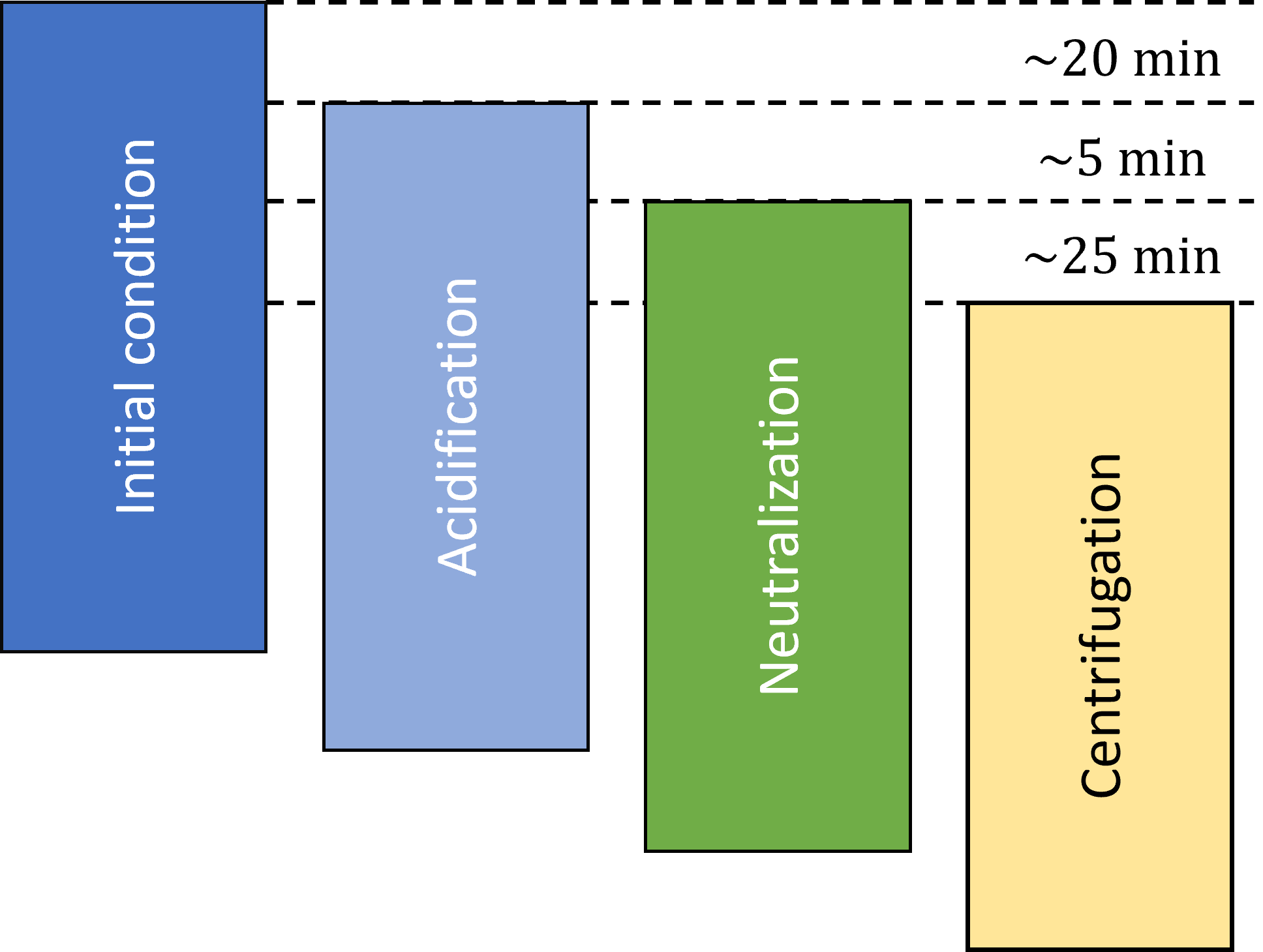}}
\caption{Approximate time between process steps}
\label{fig: Time shift}
\end{figure}

After importing the data, an important step is to connect the different values. When the oil enters at the beginning of the process, it takes about 50 min to see the result at the end of the process. Figure \ref{fig: Time shift} shows schematically how the data of the different process steps must be shifted in order to connect the values appropriately. This shifting can be done using pandas shift function \cite{Thepandasdevelopmentteam.2023}. The times shown between process steps are only an estimate. More exact values must either be measured in production or can possibly be determined using the process data.

By linking the process data this way, the first 50 minutes and the last 50 minutes of each batch are eliminated. If there occur any unwanted consequences, this approach must be reconsidered.

After importing and connecting the data, data cleansing must be performed. In the first instance, if there is not a significant number of missing values, a simple reduction should be carried out, so rows with missing data are deleted.

When looking at the aim of the analytics, to reduce the percentage of oil in the sediment, another useful step is to delete all the datasets, where the percentage of oil in the sediment is above a certain value.
Since the percentage of oil in the sediment should be below two percent, a cut at three percent would be a reasonable first step. Process conditions where the percentage is higher are not interesting in this case. \newline

The volume flow rates of acid, base and water alone have low informative value. They must always be seen in relation to the volume flow of the crude oil. For this reason, the variables in question are set as a function of the crude oil flow rate. The resulting adjustment of the predictors is shown in Table \ref{tab: Predictors and target variables - revised version}.

A simple and widely used method to deal with extreme values in normally distributed data is the use of standard deviation (\(\sigma\)), where every value (\(x\)) outside of a certain range around the mean (\(\mu\)) is sorted out, as in:
\begin{equation}
\mu - n \cdot \sigma \leq x \leq \mu + 
n \cdot \sigma\label{eq}
\end{equation}
Often three times (\(n\)) the standard deviation is used, but the exact range must be specified depending on the data. This method can likely be used for all of the parameters, since they are expected to fluctuate around a mean value.
In order to determine whether the use of the standard deviation really makes sense for all parameters, an evaluation based on the process data must be carried out \cite{Maimon.2005}.

{\renewcommand{\arraystretch}{1.2} 
\begin{table}[htbp]
\caption{\emph{Selected predictors and target variable for the training of the machine learning model}}
\label{tab: Predictors and target variables - revised version}
\centering
	\begin{tabular}{ | p {5.2cm} | p {1cm} | p {1cm} | }
		\hline
		\textbf{Description} & \textbf{Unit} & \textbf{Freq.} \\
		\hline $Acidity\, of\, crude\, Oil$ & $\%$ & $4h$ \\ 
		\hline $Relative\, acid\, flow = \frac{Flow\, of\, acid}{Flow\, of\, crude\, oil} $ & $l/kg$ & $10\, sec$ \\ 
		\hline $Relative\, base\, flow = \frac{Flow\, of\, base}{Flow\, of crude\, oil}$ & $l/kg$ & $10\, sec$ \\ 
		\hline $Relative\, water\, flow = \frac{Flow\, of\, water}{Flow\, of\, crude\, oil}$ & $l/kg$ & $10\, sec$ \\
		\hline $Oil\, content\, in\, sediment$ & $\%$ & $10\, sec$ \\ \cline{1-3}
    \end{tabular}
\end{table}
}

The data are already available in numerical form and prepared for training the analytical model, so no further processing is required.

\subsection{Data Transformation}

The data are first divided into features and targets, with the percentage of oil in the sediment as the target.
After that, the data must be divided into a test group and a training group. A typical distribution is 70/30, which means that 70 \% of the data are used for training and 30 \% for subsequent validation. In this case, because there is a high amount of available data, the test set can further be divided into two subsequent sets. One of them can be used for optimization and validation and the other one can be used for comparing the model to other approaches \cite{Kahloot.2021}.
To obtain representative samples for both groups, the data will be split while it is ordered in time, so no shuffle is carried out. In practice, the model also has to simulate the process based on data from the past. The training set itself can be split randomly into two groups. All these steps can be done using sklearn.

\subsection{Data Mining}

A regression tree is a supervised machine learning algorithm that has a continuous variable as the output \cite{Grus.2015}.
In principle, it works by splitting a dataset into two or more groups at each stage based on a feature (predictor), starting from the root node. The data set is then further divided until only a single data point remains or the creation of further branches is prevented. The value of the leaf node and thus the output is then equal to the value of the data point or the average of all data points related to the target variable.

When training a regression tree, one of the most important decisions is which principle to use to determine which features and values are used in which order to split the data set.
Widely used algorithms that are directly supported by sklearn are ID3, C4.5, C5.0 and CART \cite{scikitlearn.2022}. CART is the only algorithm out of these, that supports numeric target variables and it is also very well suited for the problem at hand. A characteristic of this algorithm is that it produces only binary regression trees, so each node always produces two other nodes \cite{Bashir.2014, JavedMehediShamrat.2022}.

In addition to the algorithm itself, supplementary parameters must be chosen. One of the most important is the splitting criterion, which influences the order in which the dataset is split and how it is split. Here, the mean squared error should be used in a first approach. This means that the square of the distances between the target variable and the mean value of the target variable is calculated. When creating the regression tree, the data set is then divided in such a way that the sum of the mean squared values is as small as possible after every step.
Other parameters like the stopping criterion have to be specified in the optimization of the model. The different approaches can then be compared using one the test sets prepared before by calculating the mean squared error for the result of each iteration. The smaller the mean squared error, the better \cite{scikitlearn.2022b}.

\subsection{Analytical Model}

\begin{figure}[htbp]
\centerline{\includegraphics[width = 0.98 \columnwidth]
{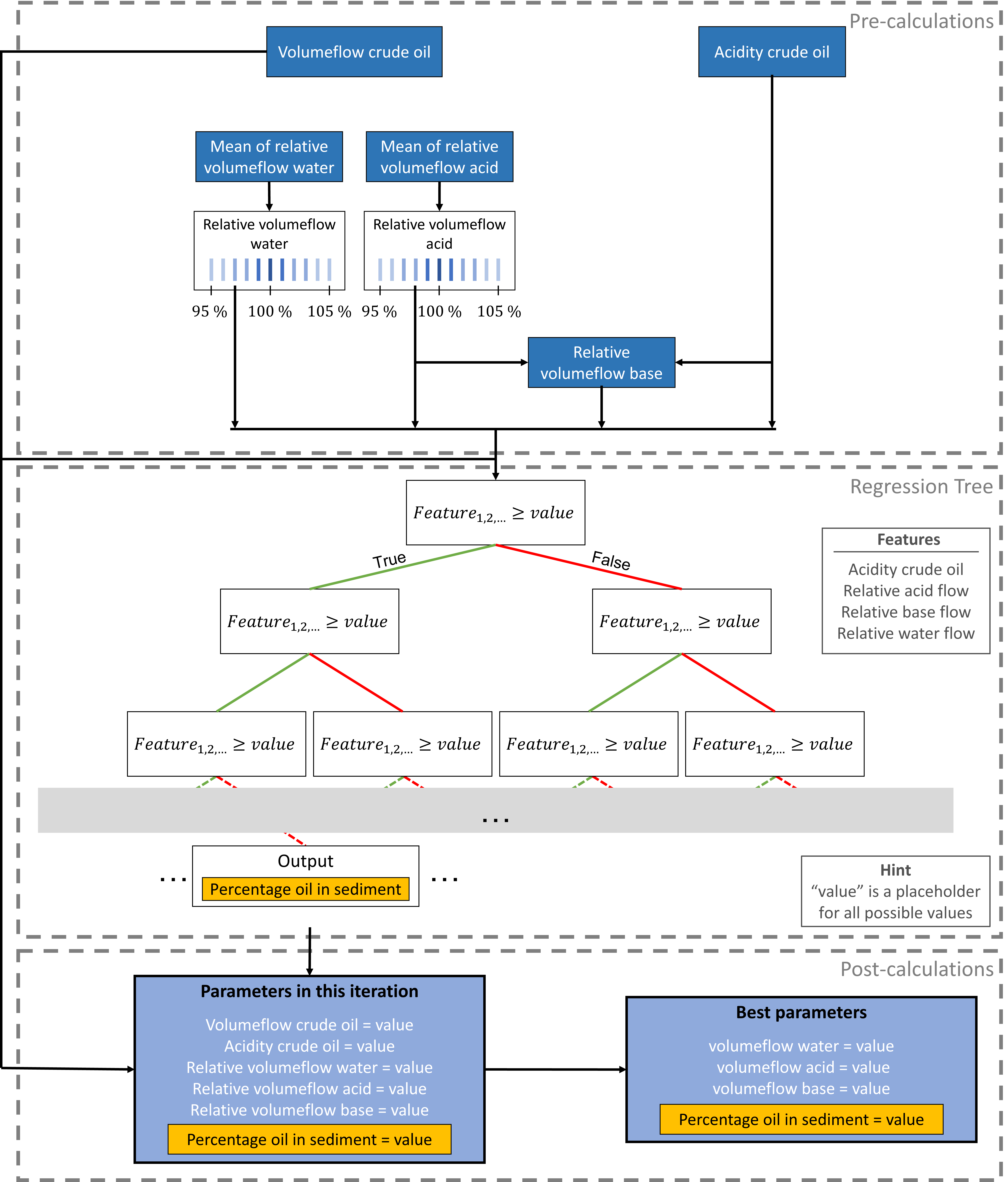}}
\caption{Schematic representation of the proposed analytical model}
\label{fig: Analytic}
\end{figure}

In order to fulfill the actual task with the help of the created regression tree, to calculate suitable process parameters for the reduction of the oil content in the sediment, pre-and post-calculations are necessary. A schematic representation of the specified analytical model can be seen in Figure \ref{fig: Analytic}.

In the pre-calculations, the means of the relative volume flows of water and acid are calculated.
The relative volume flow of water and acid are then varied in steps in a certain range around their mean value. In the schematic example in Figure \ref{fig: Analytic} the range is ± 5 \% with steps of 1 \%. In total, the process is run through 11 * 11 = 121 times. The exact range and step size must be determined with the help of test runs.

The relative volume flow of base is calculated from the relative volume flow of acid and the acidity of the crude oil. The relationship between these parameters is expected to be linear and can therefore be calculated by multiplying them with a determined factor. In this way, it should always be possible to achieve a constant, low pH of the centrifuged oil. It is expected that this or these factors will have to be calculated once, e.g. by linear regression. If this approach does not prove successful in practice, the present regression tree must either be extended to include the pH as a target variable (multi-target regression tree), or the pH and the volume flow of base must be determined in a downstream decision tree.

In the post-calculation the parameters with the lowest resulting percentage of oil in the sediment will be set as the output of the analytical model.

In the current specification, it is proposed that the analytics only run once every four hours, whenever a new value for the acidity is available. In the meantime, the calculated volume flows of acid, base and water are adjusted by multiplying the values with a factor representing the change of the crude oil flow. This reduces the computational load of the analytical model and supports the ability to be run on low cost hardware.

\section{Evaluation and Optimization}

The evaluation includes both the validation of the analytical model as a component, as well as the validation of the entire system.
In the first part of the evaluation, the functionality of the regression tree is assessed. This is done using the already prepared test data. The mean squared error and the mean error are proposed as a first criteria.
In the second step, it must be evaluated how well the calculation of the optimal process parameters work. How stable do the calculations behave in relation to small changes over the time of the process, do the values jump strongly, what resolution is reached?
To answer these questions, the parameters to be optimized must be calculated for a longer period of time. The results are then visualized and examined to gain knowledge about the analytical models behaviour.\\

In addition to the tests of the analytical model, the integration and communication capabilities of the analytics within the automated and digitalized infrastructure of the company have to be evaluated. In a first step, it is proposed to simply test these functions using sample tests, so by sending data to the system and receiving the results.
To test the functionality in a long term and the overall performance of the analytics, it is used in production as an aid for determining optimal process parameters.
More specifically, the results of the analytics, i.e. the recommended process parameters, are transmitted to the plant operators via an interface every ten seconds. The analytics thus initially serve as a criterion for the decision-making process, the result of which is still taken by the process operators.
This allows the process operators to familiarize themselves and identify problems. Optimization of the analytics as a whole are then assessed in a reaction to these observations.

\section{Conclusion}

In the process of oil production from soybeans, impurities in the crude oil must be removed. At the end of the process, in addition to the actual oil, a residue, so-called sediment, remains which can no longer be used. This paper proposes an analytics with an analytical model that calculates optimal parameters for the volume flow of acid, base and water, to achieve the lowest possible oil content in the sediment and thus to optimize the economic efficiency of the process.

The core of the model is a regression tree, which is based on the data measured in the process. In order to be able to use this data for the model, the authors proposed to transfer the available data from the database and spreadsheets into a central database. In preparation for the training of the analytical model, a selection of parameters to be used was carried out and a way to prepare the raw data for training was shown.
In addition to the specification of the analytical model, the anlytics was positioned in RAMI 4.0 and the interactions of the analytics with other components within the automated and digitalized infrastructure of the company were shown.
Care was taken to ensure that the resulting analytics can be implemented in practice using low-cost technology and without the need for expensive software.

Preliminary results obtained in an ongoing project implementing the analytics and training the regression tree, using real process data, indicate promising results.

\section*{Acknowledgments}
This work has been founded by the German Federal Ministry of Education and Research (BMBF) through the German-Argentinian University Center (DAHZ) Bi-National Master Program in Industrial Informatics and Cyber-Physical Systems (Grant agreement N0.57530192). The authors wish to acknowledge the DAHZ for their support.

\bibliographystyle{IEEEtran}
\bibliography{literature}

\end{document}